\documentclass[useAMS,usenatbib]{mn2e}
\usepackage{epsfig}

\def\gtsima{$\; \buildrel > \over \sim \;$}
\def\ltsima{$\; \buildrel < \over \sim \;$}
\def\ga{\lower.5ex\hbox{\gtsima}}
\def\lsim{\lower.5ex\hbox{\ltsima}}
\def\kms{km s$^{-1}$}

\def\Lya{Ly$\alpha$~}

\newcommand{\CIV}{\mbox{C\,{\sc iv}}}
\newcommand{\CV}{\mbox{C\,{\sc v}}}

\newcommand{\MgII}{\mbox{Mg\,{\sc ii}}}

\newcommand{\SiIV}{\mbox{Si\,{\sc iv}}}

\newcommand{\FeII}{\mbox{Fe\,{\sc ii}}}

\newcommand{\HeII}{\mbox{He\,{\sc ii}}}

\title[]{The rise of the \CIV\ mass density at
  $z<2.5$\thanks{Based on observations collected at 
  the  European  Southern  Observatory  Very  Large  Telescope,  Cerro
  Paranal,  Chile -- Programs  166.A-0106(A), 65.O-0296(A) and  during
  commissioning  and science verification of UVES }}
\author[V. D'Odorico, F. Calura, S. Cristiani, M. Viel]{Valentina D'Odorico$^{1}$\thanks{E-mail:
dodorico@oats.inaf.it}, Francesco Calura$^{1}$, Stefano Cristiani$^{1,2}$, Matteo Viel$^{1,2}$ \\
$^1$ INAF-OATS, Via Tiepolo 11, 34143 Trieste, Italy \\
$^2$ INFN/National Institute of Nuclear Physics, via Valerio 2, 34127 Trieste, Italy }

\begin{document}

\date{}

\pagerange{\pageref{firstpage}--\pageref{lastpage}} \pubyear{2010}

\maketitle

\label{firstpage}

\begin{abstract}
The cosmic evolution of the metal content of the intergalactic
medium puts stringent  constraints on the properties of
galactic outflows and on the nature of UV background. 
In this paper, we present a new measure of the redshift evolution of
the mass density of \CIV, 
$\Omega_{\rm CIV}$, in the interval $1.5 \la z \la 
4$ based on a sample of  more than 1500 \CIV\ lines with column
densities $10^{12} \la N($\CIV$) \la 10^{15}$ cm$^{-2}$. This
sample more than doubles the absorption redshift path covered in the
range $z<2.5$ by previous samples. The result shows a significant
increase of $\Omega_{\rm CIV}$ towards the lower redshifts 
at variance with the previously pictured constant behaviour. 
\end{abstract}

\begin{keywords}
intergalactic medium, quasars: absorption lines, cosmology:
observations
\end{keywords}

\section{Introduction}

The cosmological mass density of \CIV, $\Omega_{\rm CIV}$, observed as
a function of redshift is a fundamental quantity closely related to
the metal enrichment of the intergalactic medium (IGM). 
Its apparent lack of evolution  in the redshift interval
$z\simeq [1.5,5]$ \citep{songaila01,pettini03,BSR03} is puzzling since
both the physical conditions of the IGM and the properties of the
ionizing background are thought to evolve between these epochs.  

Remarkable efforts have been spent in recent years to extend the
measure of  $\Omega_{\rm CIV}$ to redshift larger than 5
\citep{ryanweber06,simcoe06} where a decrease of the star formation
rate density is observed \citep{bunker}. If $\Omega_{\rm CIV}$ is dominated by the
metals produced in situ by the observed star forming galaxies, we
would expect a decrease of its value at those redshifts. Vice versa,
the value of $\Omega_{\rm CIV}$ could remain constant if it reflects the
metallicity of a diffuse medium pre-enriched at
very high redshift.  It should be noted, however, that this is a simplified scenario
since, as redshift increases, the observed \CIV\ absorptions likely
trace gas in structures of decreasing over-density and also the
ionizing spectrum evolves in shape and intensity. As a consequence, the behaviour of
$\Omega_{\rm CIV}$ could be different from that of $\Omega_{\rm C}$
and of the mean IGM metallicity \citep[see e.g.][]{schaye03}.  

The most recent measurements of \CIV\ absorptions in
spectra of QSOs at $z\sim6$ seem to indicate a downturn in the
\CIV\ mass density at $z>5$ \citep{becker09,ryanweber09}, though based
only on 3 detected \CIV\ lines.  

At redshift $z\la 4.5$, a fundamental measurement of $\Omega_{\rm
  CIV}$  has been carried out by \citet[][S01]{songaila01}. However,
the redshift interval $1.5<z<2$ is poorly sampled by the considered
QSO spectra. A more uniform redshift coverage is provided by the
sample of \citet{BSR03} although with fewer QSO spectra. Both 
analysis are consistent with a constant behaviour of $\Omega_{\rm
  CIV}$ in the range $[1.5,4.5]$. 
At $z < 1$, recent results based on HST UV data \citep{cooksey}
give $\Omega_{\rm CIV} = (6\pm1) \times 10^{-8}$ corresponding to a
$2.8\pm0.5$ increase over the $1.5 < z < 5$ values.  
 
In this paper, we present a new measurement of  $\Omega_{\rm CIV}$ in the
redshift range $[1.5,4]$ based on a sample of 25 high resolution, high
signal-to-noise QSO spectra plus an additional sample of 8 QSO spectra
from the literature. 

The rest of the paper is organized as follows. The data are presented
in \S~2. In \S~3 the analysis is carried out with the computation of
$\Omega_{\rm CIV}$. The results are discussed in \S~4.   
Throughout this paper, we assume $\Omega_{\rm m} = 0.26$,
$\Omega_{\Lambda} = 0.74$ and $h \equiv H_0/(100 {\rm km\ s}^{-1} {\rm
  Mpc}^{-1}) =0.72$.           
     
\section{Observational data sample}

%

The core of our sample is formed by the high resolution, high
signal-to-noise QSO spectra already described
in \citet{saitta} and \citet{dodorico}. Most of them were obtained
with the Ultraviolet and Visual Echelle Spectrograph (UVES) \citep{dekker} at the Kueyen unit of the
ESO VLT (Cerro Paranal,  Chile) in  the framework of  the ESO Large
Programme (LP): ``The Cosmic Evolution of the IGM''
\citep{bergeron04}.  

In this work, 3 more QSOs were added to that sample, mainly to
increase the redshift extension above $z\sim3$. UVES spectra of the
QSO: Q0055-269, PKS2000-330 and PKS1937-101, were downloaded from the
ESO Archive and reduced with the UVES pipeline following the standard
procedure. The continuum level was determined by interpolating with a
cubic spline the region of the spectrum free from evident absorption
features.   

For all the QSO in the sample (see Table~\ref{tab1}), the \CIV\ forest was defined as the
interval between the \Lya\ emission and 5000 \kms\ from the
\CIV\ emission to avoid the proximity region where most of the
intrinsic systems are found. 
The absorption features present in this wavelength interval were
identified inspecting the spectra by eye to look for the most common doublets
(\CIV, \MgII\ and \SiIV). Then, other lines were identified testing
their compatibility with the \CIV, \MgII\ and \SiIV\ redshifts. 
Finally, lines whose identity was still unknown after this
operation, were associated with metal systems detected in the
\Lya forest or recognized as part of other multiplets (e.g.,
\FeII).

The \CIV\ doublets were fitted with Voigt profiles using
the {\small LYMAN} context of the {\small MIDAS} reduction package
\citep{font:ball}.  
A minimum number of components was adopted to fit the velocity profile
in order to reach a normalized $\chi^2\sim 1$. The fit parameters for
all the detected \CIV\ lines are reported in
Table~\ref{tab:lines}\footnote{The complete tables are available only
  in electronic format.}.     
In the following, we will refer to \CIV\ {\em components} or
  simply \CIV\ {\em lines} meaning the velocity components in which
  every absorption profile 
  has been decomposed. However, in order to compare our results with
  previous works and with data at lower resolution, we will work also
  with \CIV\ {\em systems} formed by groups of components. \CIV\ {\em
    systems} were defined in the following way: for each list of
  \CIV\ components corresponding to a single QSO the velocity
  separations among all the lines have been computed and sorted in
  ascending order. If the smallest separation is less than $dv_{\rm
    min} = 50$ \kms
  (corresponding to the velocity separation adopted by S01) the two
  lines are merged into a new line with column
  density equal to the sum of the column densities,
  and redshift equal to the average of the redshifts weighted with the 
column densities  of the components. The velocity separation are then
computed again and the procedure is iterated till the smallest
separation becomes larger than $dv_{\rm min}$. 

Our sample consists of 1023 \CIV\ velocity components with
column densities $10^{12} \la N($\CIV$) \la 10^{15}$ cm$^{-2}$ and of
508 \CIV\ systems in the same column density range.

The \CIV\ absorptions in the spectra of 19 of the QSOs forming our
sample were already identified and fitted by \citet{scannapieco} using
the software package {\small
  VPFIT}\footnote{http://www.ast.cam.ac.uk/$\sim$rfc/vpfit.html}.  We
refer to that paper for a careful 
analysis of the clustering properties of \CIV\ lines and a comparison
with the properties of \SiIV, \MgII\ and \FeII\ lines. 
Unfortunately, the authors did not publish the individual Voigt
parameters of their fitting. 

\begin{table}
\begin{center}
\caption{Relevant properties of the QSOs forming the total sample. See text for
  further details. }
\begin{minipage}{70mm}
\label{tab1}
\begin{tabular}{@{} l l c }
\hline  
QSO & $z_{\rm em}$ & $\Delta z_{\rm C IV}$  \\
\hline
HE 1341-1020 & 2.142 & 1.467-2.090 \\
Q0122-380   & 2.2004 & 1.513-2.147 \\
PKS 1448-232 & 2.224 & 1.531-2.171 \\
PKS 0237-23  & 2.233 & 1.538-2.179 \\
J2233-606   & 2.248 & 1.550-2.194 \\
HE 0001-2340 & 2.265 & 1.564-2.211 \\
HS 1626+6433$^a$ & 2.32  & 1.607-2.265 \\
HE 1122-1648 & 2.40  & 1.665-2.344 \\
Q0109-3518   & 2.4057 & 1.674-2.349 \\
HE 2217-2818  & 2.414 & 1.681-2.357 \\
Q0329-385    & 2.435 & 1.697-2.378 \\
HE 1158-1843  & 2.448 & 1.707-2.391\\
HE 1347-2457  & 2.5986 & 1.826-2.539 \\
Q1442+2931$^a$   & 2.661 & 1.875-2.600 \\
Q0453-423    & 2.669  & 1.881-2.608 \\
PKS 0329-255  & 2.696  & 1.902-2.635 \\
HE 0151-4326  & 2.763  & 1.955-2.701 \\
Q0002-422    & 2.769  & 1.959-2.707 \\
HE 2347-4342  & 2.880  & 2.067-2.816 \\
SBS 1107+487$^a$  & 2.966 & 2.114-2.900 \\
HS 1946+7658  & 3.058  & 2.181-2.991 \\
HE 0940-1050  & 3.0932 & 2.214-3.025 \\
Q0420-388    & 3.1257 & 2.239-3.057 \\
S4 0636+68$^a$ & 3.175  & 2.278-3.106 \\
SBS 1425+606$^a$ & 3.199 & 2.297-3.129 \\
PKS 2126-158 & 3.292   & 2.370-3.221 \\
B1422+231   & 3.623   & 2.630-3.546 \\
Q0055-269   & 3.66    & 2.659-3.583\\
PKS 2000-330 & 3.783   & 2.756-3.704 \\
PKS 1937-101 & 3.787   & 2.770-3.400 \\
PSS J1646+5514$^a$ & 4.059 & 2.972-3.975 \\
PSS J1057+4555$^a$ & 4.131 & 3.029-4.046 \\
BR 2237-0607$^a$   & 4.559 & 3.365-4.467 \\
\hline
\end{tabular}

\footnotesize{$^a$ QSOs from BSR03}
\end{minipage}
\end{center}
\end{table}

\begin{table}
\begin{center}
\caption{\CIV\ absorption lines: HE1341-1020 ($z_{\rm em} = 2.142$)}
\label{tab:lines}
\begin{tabular}{@{} c c c }
\hline  
$z$ & $b$ & $\log N$(\CIV$)$  \\
  & km/s  & \\
\hline
 1.699561 & $15.4\pm  0.7$  & $ 13.44 \pm 0.03 $  \\      
 1.699690 & $ 5.7\pm  0.3$  & $ 13.68 \pm 0.02 $  \\     
 1.699818 & $10.3\pm  0.6$  & $ 13.54 \pm 0.03 $  \\     
 1.700051 & $ 7.3\pm  0.6$  & $ 12.87 \pm 0.04 $  \\     
 1.700284 & $18.9\pm  1.6$  & $ 13.07 \pm 0.03 $  \\     
 1.700996 & $13.9\pm  1.2$  & $ 13.04 \pm 0.07 $  \\     
 1.700812 & $29.6\pm  2.5$  & $ 13.00 \pm 0.08 $  \\     
 1.701957 & $ 6.9\pm  0.3$  & $ 12.71 \pm 0.01 $  \\     
 1.703648 & $ 9.5\pm  0.4$  & $ 12.78 \pm 0.01 $  \\     
 1.854894 & $10.2\pm  1.0$  & $ 12.39 \pm 0.03 $  \\     
 1.910582 & $ 7.5\pm  0.5$  & $ 12.54 \pm 0.02 $  \\     
 1.910986 & $13.7\pm  1.8$  & $ 12.33 \pm 0.05 $  \\     
 1.911525 & $ 3.9\pm  1.5$  & $ 11.88 \pm 0.07 $  \\     
 1.914971 & $ 9.6\pm  1.7$  & $ 12.16 \pm 0.08 $  \\     
 1.915286 & $16.2\pm  1.8$  & $ 12.99 \pm 0.04 $  \\     
 1.915507 & $ 9.2\pm  0.8$  & $ 12.75 \pm 0.06 $  \\     
 1.998137 & $ 6.9\pm  0.7$  & $ 12.20 \pm 0.03 $  \\     
 2.041423 & $ 9.1\pm  0.4$  & $ 13.06 \pm 0.03 $  \\     
 2.041588 & $ 9.9\pm  0.9$  & $ 12.80 \pm 0.06 $  \\     
 2.084978 & $10.5\pm  0.3$  & $ 12.88 \pm 0.01 $  \\     
\hline
\end{tabular}

\end{center}
\end{table}


\subsection{Additional data sample}

In order to further increase the number of \CIV\ lines and to extend
the sample to higher redshift, we have considered the \CIV\ lines
fitted in 9 QSO spectra observed with the High Resolution Echelle
Spectrometer (HIRES) at Keck at a resolution and
signal-to-noise ratio similar to those of our spectra and reported in
\citet[][BSR03]{BSR03}.    
The fit with Voigt profiles was carried out by the authors with {\small
  VPFIT}. 
The main difference between {\small LYMAN} and {\small VPFIT} is that
the number of components fitted to a given velocity profile is, in
general, larger using the latter \citep[see also the discussion
in ][]{saitta}. 
This is seen also in the present case, in particular from the comparison of
the \CIV\ lines detected 
in the spectrum of the QSO B1422+231, which is the only object in
common between the two samples.   
We find that in all cases the number of components found with {\small
  VPFIT} is larger or equal to that found with {\small LYMAN}.   
However, when the total column density of each absorption system is
considered the
difference between the two fitting procedures becomes negligible. 



\begin{figure}
\begin{center}
\includegraphics[width=9cm]{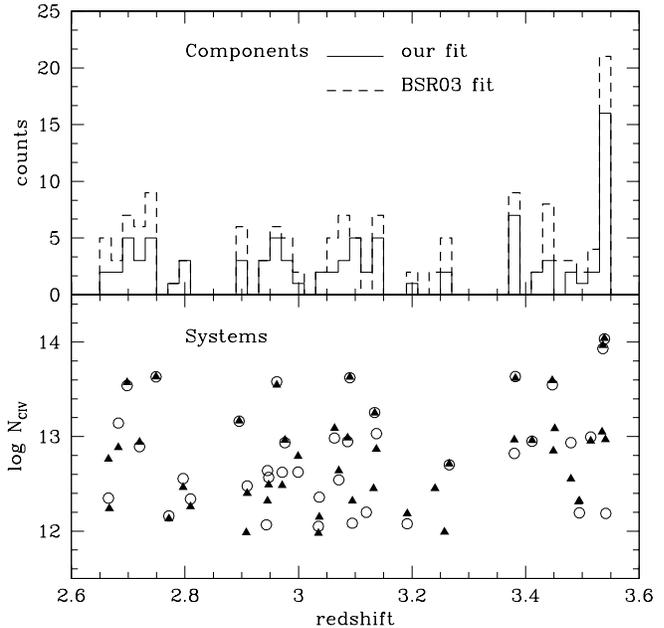}
\caption{{\it Upper panel:} Number of components of the
  \CIV\ absorption systems in the spectrum of B1422+231 as fitted by
  us (solid line) and by BSR03 (dashed line). {\it Lower panel:}
  Column densities of the \CIV\ systems of B1422+231 computed by us
  (open dots) and by BSR03 (solid triangles).}  
\label{b1422_2}
\end{center}
\end{figure}

This is shown in Fig.~\ref{b1422_2} where the number of
\CIV\ components (upper panel) and the total column density of \CIV\ systems (lower panel)
obtained with the two fitting packages are compared. While the number
of components is significantly larger in the fit by BSR03 the total
column densities are most of the time in very good agreement. 


The sample of BSR03 is formed by 577 \CIV\ components and 302
\CIV\ systems in the column density range $10^{12} \la N($\CIV$) \la 10^{15}$ cm$^{-2}$.  
In the following, we will refer to our sample of \CIV\ lines as {\em Sample
A} and to the BSR03 sample as {\em Sample B} (excluding the lines of
B1422+231, which are already in Sample A). 
The {\em total sample} is the sum of Sample A and Sample B. 
All the QSOs forming this sample are reported in Table~\ref{tab1} with
their emission redshift and the redshift range covered by the
\CIV\ forest.

\begin{figure}
\begin{center}
\includegraphics[width=9cm]{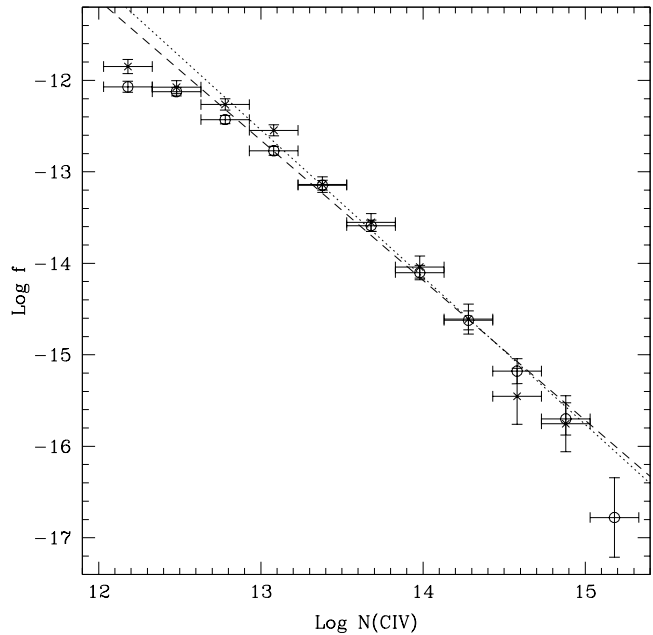}
\caption{Comparison of the column density distribution function of \CIV\ systems
  for Sample A (open dots) and Sample B (crosses) in the common redshift range $1.6 < z < 3.6$. The bin-size is
  $10^{0.3}\,N($\CIV$)$ cm$^{-2}$ and the error bars are $\pm1\,\sigma$ based
  on the number of points in each bin. The dashed and dotted lines are
  power laws of the form $f(N) = BN^{-\alpha}$ with index
  $\alpha=1.53$ and 1.71, respectively (see text).}  
\label{comp_cddf}
\end{center}
\end{figure}


\section{Data Analysis}

\subsection{ The \CIV\ column density distribution function, $f(N)$}

$f(N)$ is extremely sensitive to the velocity
decomposition of absorption features, since it is defined as the number
of lines per unit column density and per unit
redshift absorption path, $dX$ \citep{tytler87}. 
In the assumed cosmology, the redshift absorption path is: 

\begin{equation}
dX \equiv (1+z)^2 [ \Omega_{\rm m}(1+z)^3 + \Omega_{\Lambda}]^{-1/2}
dz.
\end{equation}  


In order to compare with previous works, $f(N)$ has been computed for
the \CIV\ systems in samples A and B in the common
redshift interval, $1.6<z<3.6$ (see Fig.~\ref{comp_cddf}). 
%
The excess of low column density systems in BSR03 
is due to their over-decomposition of the \CIV\ velocity profiles with
respect to our fit. 
%
A maximum likelihood fit to the data with column densities $\log
N($\CIV$) \ge 13$ (binned in Fig.~\ref{comp_cddf} for display
purposes only) to a power law of the form $f(N) = B\,N^{-\alpha}$, 
gives an index  $\alpha=1.71\pm0.07$ for sample A and $\alpha=1.8\pm0.1$ for
sample B. Both are in agreement with the result by S01 based on
\CIV\ systems defined in the same way.  
These are steeper than the \citet{ellison} fit of $1.44\pm0.05$ based
on the very high SNR spectrum of B1422+231 and extending the power law
down to $\log N($\CIV$) \sim 12.3$. 

\begin{table}
\begin{center}
\caption{Results of incompleteness tests}
\label{complet}
\begin{tabular}{@{} c c  }
\hline  
$\log N($\CIV$)$ & Fraction of detected \CIV  \\
\hline
12.00  & 0.60 \\
12.30  & 0.82 \\
12.60  & 0.97 \\
12.78  & 0.97 \\
13.00  & 1.00 \\
\hline
\end{tabular}
\end{center}
\end{table}

In order to estimate the incompleteness of our data for $\log
N($\CIV$) \le 13$ we have performed the following simulations. Four
fake \CIV\ doublets have been generated for each spectrum in our sample, with
column densities $12 \le N($\CIV$) \le 13$  and redshift chosen
randomly from the redshift ranges reported in Table~\ref{tab1}. The
$b$-values of the fake 
\CIV\ lines were drawn at random from the observed distribution of
$b$-values in our data. These fake \CIV\ doublets were then added to
the real spectra and searched for by the member of our team who had
previously identified the real \CIV\ lines. Given the visual, rather
than automatic, character of our searches, only two such trials were
performed. The results of these tests are collected in
Table~\ref{complet}. 

We found that we could recover essentially all \CIV\ doublets as long
as $\log N($\CIV$) \ge 12.6$. For $\log N($\CIV$) =12.3$ a correction
factor of 1.2 has been determined (31/38 fake \CIV\ systems detected
in the incompleteness tests). In the lowest column density bin, $\log
N($\CIV$) =12$, 24 over 40 \CIV\ lines have been identified,
corresponding to a correction factor of 1.7. The average Doppler
parameter of the undetected \CIV\ lines is $\langle b \rangle \sim 17.5$
\kms, while detected lines have $\langle b \rangle \sim 9$ \kms. 

Extending the range of fitted data to our completeness limit, $\log
N($\CIV$) \ge 12.6$, the maximum likelihood indexes become
$\alpha=1.53\pm0.04$ for sample A and $\alpha=1.61\pm0.07$ for sample
B, in agreement with the result by \citet{ellison}. 
If, the correction factors reported above are applied to the points of
Sample A,  their values increase slightly to coincide with the values in Sample
B shown in Fig.~\ref{comp_cddf}. The corrected value of the lowest
column density bin is still too low to be fitted with the previously
derived power laws.

\begin{figure}
\begin{center}
\includegraphics[width=9cm]{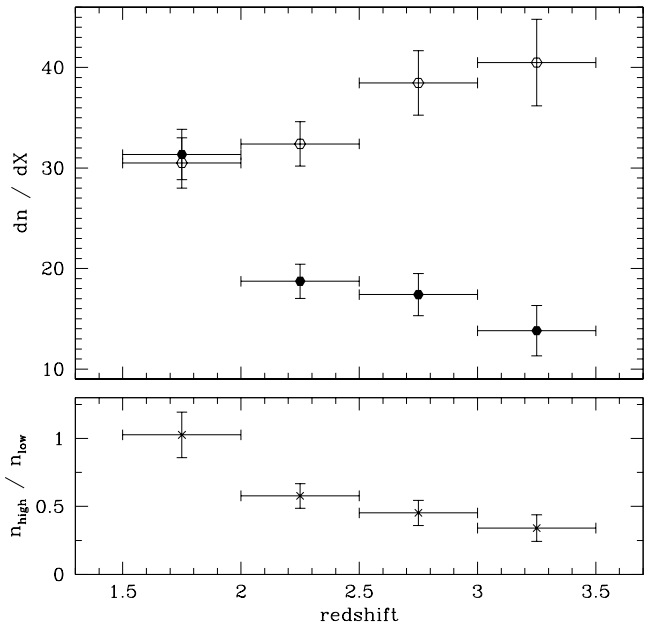}
\caption{{\it Upper panel:} Redshift evolution of the number density
  of the \CIV\ lines 
  in Sample A divided into low column density (open dots,
  $12 \leq \log N($\CIV$) \leq 13$) and high column density (solid
  dots,  $13 < \log N($\CIV$) \leq 15$) absorptions. {\it Lower
    panel:} Redshift evolution of the ratio between the number of high
  and low column density \CIV\ lines. 
}   
\label{ndens}
\end{center}
\end{figure}

\subsection{ The number density of \CIV\ lines, $dn/dX$}

The evolution with redshift of $dn/dX$ computed for \CIV\ lines in Sample A
is reported in the upper panel of Fig.~\ref{ndens}. In particular, we
have split our sample into weak and strong lines characterized by column
densities $12 \leq \log N($\CIV$) \leq 13$ and $13 < \log N($\CIV$)
\leq 15$, respectively. The two populations show different
behaviours, with the weak lines being consistent with no evolution and
the strong ones significantly decreasing in number towards higher
redshifts. The latter trend confirms the result by \citet{steidel}.
Furthermore, the ratio of the number of strong and weak  
lines shows a steady decrease with redshift. 

\begin{figure}
\begin{center}
\includegraphics[width=9cm]{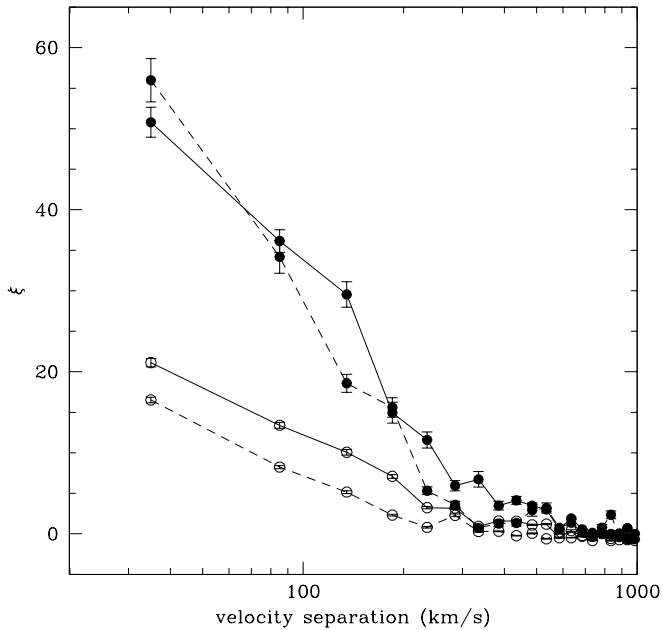}
\caption{Two-point correlation function of the \CIV\ lines 
  in Sample A divided into low column density (open dots,
  $12 \leq \log N($\CIV$) \leq 13$) and high column density (solid dots,
  $13 < \log N($\CIV$) \leq 15$) absorptions. Solid lines trace 
  the result for the redshift bin $1.5 \leq z \leq 2.3$ while dashed
  lines refer to the bin $2.3 < z \leq 3.5$.
}   
\label{tpcf}
\end{center}
\end{figure}

\subsection{ The two-point correlation function of \CIV\ lines, $\xi(v,v+\Delta v)$} 

We computed $\xi(v,v+\Delta v)$ for the weak and strong
\CIV\ components and for two redshift intervals $1.5 < z < 2.3$ and
$2.3 < z < 3.5$ (see Fig.~\ref{tpcf}). 
The obtained two-point correlation functions
show a characteristic clustering scale $r_0$ (in redshift space) for
the 
sample at high column densities which is about $1.5-2$ times larger  than for
the low column density lines ($\sim 850$  vs. 500 km s$^{-1}$ in the
high redshift bin and 700 vs. 314 km s$^{-1}$ for the low redshift
bin, with $1\,\sigma$ errors of about 150 km s$^{-1}$). Converting
these scales into comoving 
Mpc $h^{-1}$ at a given redshift, it appears that the low column density
sample has clustering properties similar 
to those of Lyman-break galaxies (LBGs) at $z\sim 3$
\citep[e.g.][]{porciani}, while large column density \CIV\ systems are
more strongly clustered and possibly  sample denser environment and
more massive objects. \citet{adelberger} reached similar
conclusions from the analysis of a large sample of LBGs observed in
the fields of 23 high redshift QSOs.  



\subsection{The redshift evolution of the mass density of \CIV}

We used our sample of absorption lines also to compute the mass
density of \CIV\ as a fraction of the 
critical density today:

\begin{equation}
\label{omega}
\Omega_{\rm CIV} = \frac{H_0\, m_{\rm CIV}}{c\, \rho_{\rm crit}} \int N
f(N) dN, 
\end{equation}
where $H_0 = 100\, h$ \kms Mpc$^{-1}$ is the Hubble constant, $ m_{\rm
  CIV}$ is the mass of a \CIV\ ion, $c$ is the speed of light,
$\rho_{\rm crit} = 1.88 \times 10^{-29} h^2$ g cm$^{-3}$ and $f(N)$ is
the Column Density Distribution Function (CDDF).  
Since $f(N)$ cannot be recovered correctly for all column densities
due to incompleteness and poor statistics, the integral in eq.~\ref{omega} can be
approximated by a sum: 

\begin{equation}
\label{omega_approx}
\Omega_{\rm CIV} = \frac{H_0\, m_{\rm CIV}}{c\, \rho_{\rm crit}}
\frac{\sum_i  N_i (\mbox{\CIV})}{\Delta X},
\end{equation}
with an associated fractional variance:

\begin{equation}
\label{omega_err}
\left( \frac{\delta \Omega_{\rm CIV}}{\Omega_{\rm CIV}} \right)^2 = \frac{\sum_i  [N_i (\mbox{\CIV})]^2}{\left[\sum_i  N_i (\mbox{\CIV})\right]^2}
\end{equation}
as proposed by \citet{storrie96}. 
Error bars have been computed also using a bootstrap technique to
build 1000 samples of QSO spectra, based on the observed sample, for
each redshift bin and determining the standard deviation of the
resulting distribution of $\Omega_{\rm CIV}$ values. 
This estimate is always slightly larger (by a factor of $\sim 1.5$ or
smaller) than the one based on
eq.~\ref{omega_err} and has been used as the reference one.  
 
The value of $\Omega_{\rm CIV}$ significantly depends on the column
density range over which the sum or the integration are carried out,
and as a consequence on the resolution and SNR of the available
spectra. To take this aspect into account, we have computed three sets
of values to be compared consistently with different data in the
literature.  

\begin{figure}
\begin{center}
\includegraphics[width=9cm]{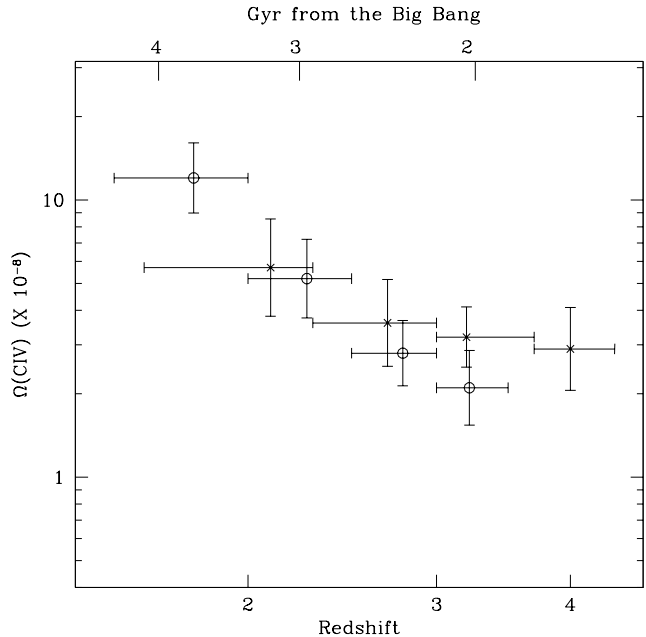}
\caption{Comparison of the $\Omega_{\rm CIV}$ determinations based on
  \CIV\ lines in Sample A (open dots) and the results by BSR03 (crosses). 
  Error bars are $1\,\sigma$. }   
\label{omegaciv_comp}
\end{center}
\end{figure}

$\Omega_{\rm CIV}$ obtained for the \CIV\ components in Sample A with
column densities $12 \leq \log N($\CIV$) \leq 15$ is compared
with the values reported in the original paper by BSR03 in
Fig.~\ref{omegaciv_comp}.   Our results confirm and strengthen the
increasing trend with decreasing redshift of $\Omega_{\rm CIV}$ which
was already present in the data by BSR03.

\begin{figure}
\begin{center}
\includegraphics[width=9cm]{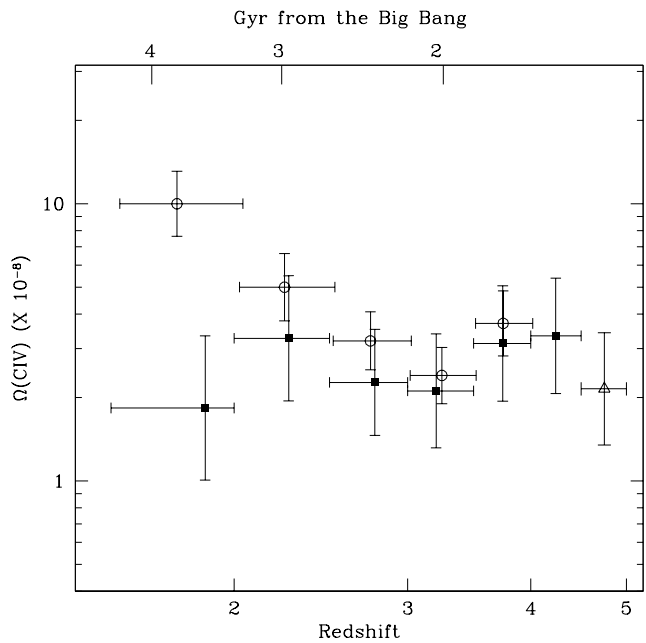}
\caption{$\Omega_{\rm CIV}$ determined for the systems in the total sample (open
  dots) with $1\,\sigma$ error bars from bootstrap method compared
  with the results by S01 (solid squares) and by \citet{pettini03}
  (empty triangle). }  
\label{omegaciv_syst}
\end{center}
\end{figure}

In order to carry out the comparison with the data set by S01, the
most used in the literature for the redshift range  $[1.5,4.5]$,
$\Omega_{\rm CIV}$ has been computed from the
\CIV\ systems\footnote{We remind the readers that \CIV\ systems have
  been built adopting a minimum velocity separation of 50 \kms\ as
  reported in S01.} of the total sample with column densities $12 \leq \log
N($\CIV$) \leq 15$. In Fig.~\ref{omegaciv_syst}  these results are
reported together with the value of S01 corrected for the assumed
cosmology (they assumed Einstein-de Sitter) and with error bars
derived from their plot and transformed to $1\,\sigma$.  
The present data sample represents an increase in the absorption
redshift path of a factor 2.5 and 2 in the redshift bins
[1.5,2.0] and [2.0,2.5] respectively, with respect to S01. Indeed,
our estimate in the lowest redshift bin is $\sim 4.5\,\sigma$ larger
than the previously accepted value, suggesting a trend of increasing
\CIV\ mass going to lower redshifts. The computed values with the
associated errors are reported in Table~\ref{tab2}. 

We have not applied corrections to the values of $\Omega_{\rm CIV}$ due to the
incompleteness of our observations for \CIV\ lines with column
densities $\la 12.6$, since we estimated them to be less than 3 \% (less than 1
\% in the lowest redshift bin). This is due to the small correction
factors determined in Section~3.1 and to the small contribution of low
column density lines to the final values. 
   
%

\begin{figure}
\begin{center}
\includegraphics[width=9cm]{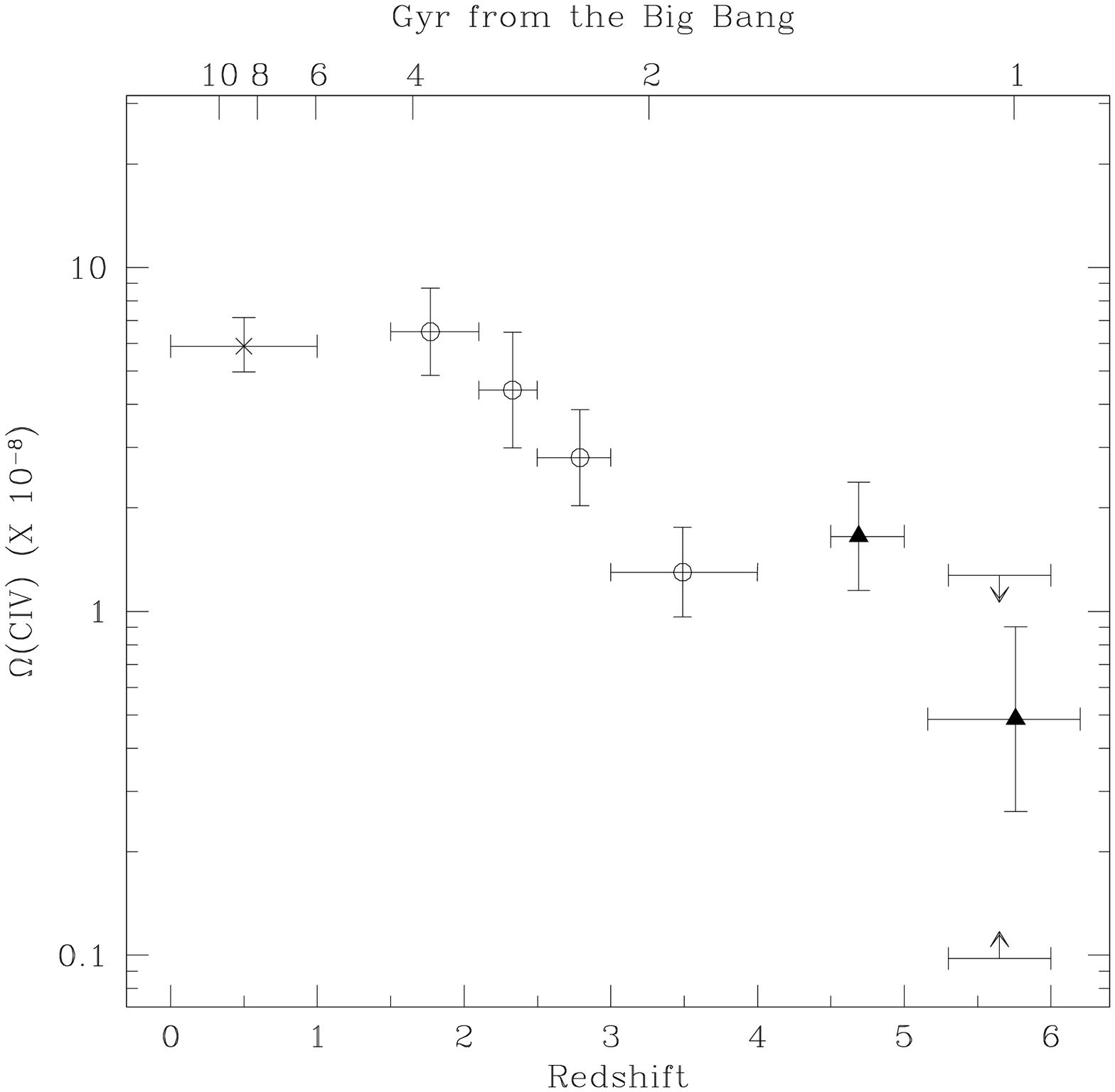}
\caption{Estimates of $\Omega_{\rm CIV}$ in the redshift range
  $z\simeq 0-6$ for \CIV\ systems with $13.8 \le \log N($\CIV$) \le 15$: {\it cross}
  \citet{cooksey}; {\it open dots} total 
  sample this work; {\it solid triangles} \citet{pettini03} and
  \citet{ryanweber09};  {\it 95 \% confidence interval} \citet{becker09}.}  
\label{omegaciv_all}
\end{center}
\end{figure}

Recently, the determinations of $\Omega_{\rm CIV}$ have been extended
to very low \citep[$z<1$, e.g.][]{cooksey} and very high
\citep[$z>5$][]{ryanweber09,becker09} redshift. Observations in 
these redshift ranges are more difficult since the \CIV\ transition
moves to the UV and to the IR region of the electromagnetic spectrum,
respectively. As a consequence, the estimates of $\Omega_{\rm CIV}$ are based on lower
resolution and lower SNR spectra limiting the detectability of
\CIV\ lines to larger column densities. In order to study the
evolution of  $\Omega_{\rm CIV}$ in the whole redshift range between
$z\sim0$ and 6, we have carried out a third computation of the values
of  $\Omega_{\rm CIV}$ for \CIV\ systems in the column density range $13.8 \leq \log
N($\CIV$) \leq 15$, changing the definition of the redshift bins in
order to have comparable absorption redshift paths covered in each one
of them. The column density lower boundary is the limit to which
  the higher (and lower) redshift surveys cited above are sensitive
  (being based on lower resolution and SNR spectra than those analysed in the present
work). The results are shown in Fig.~\ref{omegaciv_all} where the
point at $z\simeq 4.69$ is the determination by \citet{pettini03}
corrected for the considered column density range by
\citet{ryanweber09}. The plot shows clearly  the smooth growth of 
the value of $\Omega_{\rm CIV}$ from the plateau at redshifts $3-5$ to
the local value, corresponding to an increase of a factor  $\sim
5$. 
Our determinations are reported also in
Table~\ref{tab2}.

\begin{table}
\begin{center}
\caption{$\Omega_{\rm CIV}$ for the systems in the total sample
  selected in the reported column density intervals}
\label{tab2}
\begin{tabular}{c r c c c c}
\hline  
$z$ range & $\Delta$ X & lines & $\Omega_{\rm CIV}$  & $\delta
\Omega$  & $(\delta\Omega)_{\rm boot}$\\ 
 & & & ($\times 10^{-8}$) & ($\times 10^{-8}$) & ($\times 10^{-8}$) \\
\hline
\multicolumn{3}{c}{ $12 \leq \log N($\CIV$) \leq 15$} & & & \\
$1.5-2.0$ &  16.61 & 148 & 10.0 &  1.8  & 2.7 \\
$2.0-2.5$ &  28.11 & 211 & 5.0 & 1.0  & 1.4 \\
$2.5-3.0$ &  19.82 & 161 & 3.2 & 0.7  & 0.8 \\
$3.0-3.5$ &  13.91 & 132 & 2.4 & 0.4  & 0.6 \\
$3.5-4.0$ &   7.60 &  78 & 3.7 & 0.7  & 1.0 \\
\multicolumn{3}{c}{ $13.8 \leq \log N($\CIV$) \leq 15$} & & & \\
$1.5- 2.1$ &  22.56  & 41 & 6.5 &  1.3 & 1.9 \\
$2.1- 2.5$ &  22.16  & 23 & 4.4 &  1.2 & 1.7 \\
$2.5- 3.0$ &  19.85  & 19 & 2.8 &  0.8 & 0.9 \\
$3.0- 4.0$ &  21.50  & 17 & 1.3 &  0.3 & 0.4 \\
\hline
\end{tabular}
\end{center}
\end{table}


\section{Discussion}

In this paper, we have presented a new determination of the cosmological
mass density of \CIV, $\Omega_{\rm CIV}$, at $z=[1.5,4]$ based on
a large sample of high 
resolution, high signal-to-noise QSO spectra which more than doubles,
with respect to previous measurements, the covered absorption path for
$1.5 \la z \la 2.5$.
The main result of our calculation is that  $\Omega_{\rm CIV}$ is no
longer approximately constant in the considered redshift range, but
shows a steady increase from $z\sim3-5$ to $z\sim1.5-2$. On the other hand,
it appears that the \CIV\ mass density is not evolving significantly from
these redshifts down to the present epoch \citep{cooksey}.  

The value of $\Omega_{\rm CIV}$ for the column density interval $12
\leq \log N($\CIV$) \leq 15$ can be converted into the IGM carbon
content by mass, with the formula:  
%
\begin{equation}
Z_{\rm C} \simeq \frac{\Omega_{\rm CIV}}{\Omega_{\rm b}}\cdot
\frac{{\rm C}}{{\rm C\,IV}}
\end{equation}
where, $\Omega_{\rm b} = 0.0224/h^2$ \citep{pettini08} is the
contribution of baryons to the critical density and \CIV/C is
the fraction of C which is triply ionized, which depends on the assumed
ionizing background. 
Assuming the maximum relative abundance of \CIV, that is
\CIV/C~$\la 0.5$, 
and the solar carbon over hydrogen abundance by mass $Z_{\rm
  C,\sun}= 0.0029 $
\citep{asplund+05}, we obtain a lower limit at redshifts $[2.5,4.0]$ of  
%
$Z_{\rm C} \ga 1.4 \times 10^{-6} = 4.9 \times 10^{-4} Z_{\rm C,\sun}$
%
or [C/H]~$\ga -3.3$. 
At redshift $[1.5,2.0]$:
%
$Z_{\rm C} \ga 4.6 \times 10^{-6} = 1.6 \times 10^{-3} Z_{\rm C,\sun}$
%
or [C/H]~$\ga -2.8$, corresponding to an increase of a factor of $\sim
3$ with respect to the high redshift bin. 

The metallicity measured at high redshift  is in agreement with the result
obtained at $z=3$ by \citet{schaye03} with the pixel optical depth
method, in the case of over-densities above $\sim 10$.  This is 
consistent with the fact that $\Omega_{\rm CIV}$ traces mainly the
metallicity of the over-dense regions in the proximity of galaxies and
its redshift evolution is linked with that of
the strong systems.   
\citet{simcoe04} found a median metallicity [C,O/H]~$\simeq -2.8$ for
the intergalactic gas at $z\sim 2.5$ with a method based on the direct
detections of the absorption lines which takes into account also the
upper limits. Re-normalizing our result to the solar abundances and the
fraction of C triply ionized adopted by those authors  (\CIV/C~$\la
0.25$), we obtain  [C/H]~$\ga -3.0$ consistent with their result.

The physical explanation for the redshift evolution of $\Omega_{\rm
  CIV}$ can be explored with cosmological simulations.
We will devote a subsequent paper to the comparison between
observations and predictions of hydro-simulations with a detailed 
treatment of metal enrichment (Tescari et al., in preparation). 
In the following, we will briefly discuss the few predictions of
$\Omega_{\rm CIV}$ present in the literature and compare them with our
results.     

\citet[][OD06]{opp_dave06} have run cosmological hydro-simulations of galaxy
formation meant to reproduce the metallicity of the IGM, where metals
are expelled from galaxies by momentum-driven winds (whose velocity is
proportional to the dispersion velocity of the galaxies). Those winds are
very effective in transporting the gas without heating it too much. 
Their fiducial simulation reproduces consistently the star formation
rate of the Universe, the volume averaged metallicity of the IGM and
the lack of evolution of   $\Omega_{\rm CIV}$ from $z\approx 5 \rightarrow
1.5$, as observed in the data available at the time. 
On the other hand, the predicted mass density of C increases by nearly
an order of magnitude toward lower redshifts in the same redshift
interval, due to the increase in the total metallicity of the gas. The
constant behaviour of $\Omega_{\rm CIV}$ is obtained by balancing 
the increase in C abundance with  galactic winds. This form of
feedback heats the IGM causing a decrease of the \CIV\ ionization
fraction. In particular, the peak of the  \CIV\ ionization fraction
distribution as a function of over-density shifts toward larger
over-densities as redshift decreases. This is consistent with the
increase in the ratio between strong and weak \CIV\ systems at lower
redshifts shown in Fig.~\ref{ndens} and the identification of strong
systems with denser environment as deduced from the observed 2-point
correlation function (see Section~3.3). 
 
To conclude, none of the models proposed by OD06 predicts an increase
of $\Omega_{\rm CIV}$ toward lower redshifts. An improvement in the
physics of the wind, plus the introduction of the AGB feedback, results
in a slight increase of $\Omega_{\rm CIV}$ in the ranges $z < 0.5$ and
$z > 5$ leaving substantially unchanged the other values
\citep[][OD08]{opp_dave08}.


An effect which has not been taken into account up to now is the
evolution in the shape of the UV background due to the 
\HeII\ re-ionization process expected at 
redshift $\sim 3$. Naively, due to the increase in the number of free hard
photons (ionizing \CIV\ into \CV) at the end of the re-ionization
epoch, we would expect a trend opposite to what is observed: a
decrease of the amount of \CIV\ after $z\sim3$.  However, the details
of this process are still under study \citep[e.g.][]{bolton09,hm09}
and other factors should be accounted for properly: the decrease in the number
density of QSOs going towards lower redshifts and the fact that at low
$z$ the regions traced by the strong \CIV\ absorbers (driving the
evolution of  $\Omega_{\rm CIV}$) could be only mildly affected by the cosmic
UV background. 

The observed raise of the cosmic \CIV\ mass density in the redshift
range $1.5-2.5$ puts a strong constraints on the models describing 
the interplay between galaxies and their surrounding medium, suggesting
that something is still missing in the physical implementation of
galactic feedback.   


\section*{Acknowledgments}

We would like to thank Eros Vanzella for his precious help in
generating the synthetic \CIV\ absorbers. 
We are grateful to the anonymous referee, whose comments and advices allowed us
to improve significantly this paper.
This research has been partially supported by ASI Contract No. I/016/07/0
COFIS, INFN PD51 grant and  PRIN MIUR.
FC is supported by a fellowship by PRIN MIUR.

\bsp

\label{lastpage}

\end{document}